\documentclass[prd,superscriptaddress,twocolumn,nofootinbib]{revtex4-2}
\usepackage{color}
\usepackage{bm}
\usepackage{graphicx}  % needed for figures
\usepackage{dcolumn}   % needed for some tables
\usepackage{bm}        % for math
\usepackage[english]{babel}
\usepackage{booktabs,multirow,array}
\usepackage[bottom]{footmisc}
\usepackage{dblfnote}
\usepackage{mathtools}
\usepackage{amsmath}
\usepackage{physics}
\DFNalwaysdouble % for this example
\usepackage[colorlinks]{hyperref}
\definecolor{LinkColor}{rgb}{0.75, 0, 0}
\definecolor{CiteColor}{rgb}{0, 0.5, 0.5}
\definecolor{UrlColor}{rgb}{0, 0, 0.75}
\hypersetup{linkcolor=LinkColor}
\hypersetup{citecolor=CiteColor}
\hypersetup{urlcolor=UrlColor}
\begin{document}

\title{Thermodynamics of multi-horizon spacetimes}
\author{Chiranjeeb Singha}
\email{chiranjeeb.singha@saha.ac.in}
\affiliation{Theory Division, Saha Institute of Nuclear Physics, Kolkata 700064, India}


%\date{\today}
%\date{ }  %% This command  will suppress printing the date. 
%If date is required, comment out this line.
%%%%%%%%%%%%%%%%%%%%%%%%%%%%%%%%%%%%%%%%%%%%%%%%%%%%%%%%%%%%%%%%%%%%%%%%%%%%%%%%%%%%%%%%%%%%%%%%%%%
%%%%%%%%%%%%%%%%%%%%%%%%%%%%%%%%%%%%%%%%%%%%%%%%%%%%%%%%%%%%%%%%%%%%%%%%%%%%%%%%%%%%%%%%%%%%%%%%%%%

%%%%%%%%%%%%%%%%%%%%%%%%%%%%%%%%%%%%%%%%%%%%%%%%%%%%%%%%%%%%%%%%%%%%%%%%%%%%%%%%%%%%%%%%%%%%%%%%%%%
%%%%%%%%%%%%%%%%%%%%%%%%%%%%%%%%%%%%%%%%%%%%%%%%%%%%%%%%%%%%%%%%%%%%%%%%%%%%%%%%%%%%%%%%%%%%%%%%%%%
%%%%%%%%%%%%%%%%%%%%%%%%%%%%%%%%%%%%%%%%%%%%%%%%%%%%%%%%%%%%%%%%%%%%%%%%%%%%%%%%%%%%%%%%%%%%%%%%%%%

\begin{abstract}

There exist several well-established procedures for computing thermodynamics for a single horizon spacetime. However, for a spacetime with multi-horizon, the thermodynamics is not very clear. It is not fully understood whether there exists a global temperature for the
multi-horizon spacetime or not. Here we show that a global temperature can exist for Schwarzschild-de Sitter spacetime, Reissner-Nordstrom-de Sitter spacetime, and rotating BTZ black hole. This temperature does not coincide with the conventional Hawking temperature related to the outer horizon. We also show that the total entropy for these spacetimes can not be determined only by the outer horizon. The correlations between the horizons of these spacetimes determine it.

\end{abstract}
\maketitle

\section{introduction}

For a single horizon spacetime, we already have many well-established procedures for computing thermodynamics. It has been shown that
the temperature of the single horizon spacetime is proportional to the surface gravity at the horizon  \cite{hawking1975, Birrell1984quantum, book:parker,
Jacobson:2003vx, Kiefer:2002fp, Parikh:1999mf, Traschen:1999zr, 
DEWITT1975295, Davies:1974th, Padmanabhan:2009vy,
Wald1975}. 
However, for a spacetime with two or more horizons,  the thermodynamics is not very clear. It is not fully understood whether there exists a global temperature for the
multi-horizon spacetimes or not. If more than one horizons for a spacetime are present, it may contribute to the temperature of the spacetime \cite{Volovik:2021upi, Volovik:2021iim, Choudhury:2004ph, Chabab:2020xwr, Shankaranarayanan:2003ya}. Also, the total entropy of the spacetime can not be determined only by the outer horizon. The correlations between the multi horizons determine it \cite{Volovik:2021upi, Volovik:2021iim, Shankaranarayanan:2003ya}.

Recently, It has been shown that a global temperature can exist for a Reissner-Nordstrom black hole \cite{Volovik:2021upi, Volovik:2021iim}. This temperature depends only on the mass of the black hole and does not depend on the black hole charge. Also, the total entropy of a Reissner-Nordstrom black hole, a Kerr black hole, and a Kerr-Newman black hole can be determined by the correlations between inner and outer horizons \cite{Volovik:2021upi, Volovik:2021iim}.

Here, we consider the Schwarzschild-de Sitter spacetime, Reissner-Nordstrom-de Sitter spacetime, and rotating BTZ black hole.  We compute the temperature of the thermal radiation and the entropy for these spacetimes. We show that for these spacetimes, a global temperature can exist. This temperature does not coincide with the conventional Hawking temperature related to the outer horizon. We also show that the total entropy for these spacetimes can be determined by the correlations between the horizons of these spacetimes.

In Sec. \ref{Schwarzschild-de Sitter}, we compute the thermal radiation from the Schwarzschild-de Sitter spacetime using the semiclassical tunneling approach \cite{Parikh:1999mf, Srinivasan:1998ty, Volovik:1999fc, Akhmedov:2006pg, Vanzo:2011wq, Jannes:2011qp}.
We obtain a global temperature of radiation that does not coincide with the conventional Hawking temperature related to the outer cosmological horizon. We also compute the entropy by using the approach based on the method of the singular coordinate transformations \cite{Volovik:2021upi, Volovik:2021iim} for this spacetime. We show that the total entropy for this spacetime can not be determined only by the outer cosmological horizon. The correlations between the outer cosmological horizon and the inner black hole horizon determine it. This result is precisely matched with the result in the paper \cite{Shankaranarayanan:2003ya}.

In Sec. \ref{Reissner-Nordstrom-de Sitter}, we compute the thermal radiation from the Reissner-Nordstrom-de Sitter spacetime using the semiclassical tunneling approach \cite{Parikh:1999mf, Srinivasan:1998ty, Volovik:1999fc, Akhmedov:2006pg, Vanzo:2011wq, Jannes:2011qp}. We obtain a global temperature of radiation that does not coincide with the conventional Hawking temperature related to the outer cosmological horizon. We also compute the entropy by using the approach based on the method of the singular coordinate transformations \cite{Volovik:2021upi, Volovik:2021iim} for this spacetime. We show that the total entropy for this spacetime can not be determined only by the outer cosmological horizon. The correlations between the cosmological horizon, the event horizon, and the Cauchy horizon determine it.

In Sec. \ref{BTZ}, we compute the thermal radiation from the rotating BTZ black hole using the semiclassical tunneling approach \cite{Parikh:1999mf, Srinivasan:1998ty, Volovik:1999fc, Akhmedov:2006pg, Vanzo:2011wq, Jannes:2011qp}.
We obtain a global temperature of radiation that does not coincide with the conventional Hawking temperature related to the outer event horizon. We also compute the entropy by using the approach based on the method of the singular coordinate transformations \cite{Volovik:2021upi, Volovik:2021iim} for this spacetime. The total entropy for this spacetime can not be determined only by the outer event horizon. The correlations between the outer event horizon and the inner Cauchy horizon determine it.

\section{Hawking radiation from two horizons of Schwarzschild-de Sitter spacetime}\label{Schwarzschild-de Sitter}

The line element for the Schwarzschild-de Sitter spacetime in the spherically symmetric coordinate is given by \cite{Bhattacharya:2013tq, Shankaranarayanan:2003ya, PhysRevD.66.124009, Pappas:2017kam, Robson:2019yzx,Tian:2003ua},
\begin{equation}\label{eq1}
ds^2=-f(r)dt^2+f(r)^{-1} dr^2 +r^2 d \Omega^2~,
\end{equation}
where $f(r)=\left(1-\frac{2 M}{r}-\frac{r^2}{l^2}\right)$. Here $M$ is the mass of the black hole, and $l^2$ is connected with the positive cosmological constant. There is more
than one horizon present in this spacetime if and only if $0<y<1/2 M$, where $y=M^2/l^2$. The black hole horizon ($r_{h}$) is the inner horizon, and the cosmological horizon ($r_{c}$) is the outer horizon for this spacetime. The expression for these horizons are,
\begin{align}
r_{h}=\frac{2 M}{\sqrt{3 y}} \cos \frac{\pi+\varphi}{3}~,\\
 r_{c}=\frac{2 M}{\sqrt{3 y}} \cos \frac{\pi-\varphi}{3}~,
\end{align}
where $\varphi= \cos^{-1}(3 \sqrt{3} y)$. The surface gravities at the horizons are \cite{PhysRevD.57.2436, Shankaranarayanan:2003ya}, 
\begin{align}
\kappa_{h}= \alpha \left|\frac{M}{r^2_h}-\frac{r_h}{l^2}\right|~,\\
\kappa_{c}= \alpha \left|\frac{M}{r^2_c}-\frac{r_c}{l^2}\right|~,
\end{align}
where $\kappa_{h}$ and $\kappa_{c}$ are the surface gravities of the black hole horizon and the cosmological horizon, respectively. Here $\alpha= 1/\sqrt{1-(27y)^{1/3}}$. In terms of the following coordinate transformations,
\begin{equation}\label{eq6}
 d\tilde{t}\rightarrow dt \pm f dr,~~f=\frac{\sqrt{\frac{2M}{r}+\frac{r^2}{l^2}}}{\left(1-\frac{2 M}{r}-\frac{r^2}{l^2}\right)}~.
\end{equation}
we obtain Painleve-Gullstrand (PG) metric for Schwarzschild-de Sitter spacetime as,
\begin{equation}\label{eq7}
ds^2= g_{\mu \nu} dx^{\mu} dx^{\nu}=-dt^2+(dr\pm v dt)^2+r^2 d\Omega^2~.
\end{equation}
Here the $v$ is the shift velocity which is given by,
\begin{equation}
 v^2=\frac{2 M}{r}+\frac{r^2}{l^2}~.
\end{equation}

The tunneling trajectory for a massless particle can be found as,
\begin{equation}
 g^{\mu\nu}p_{\mu}p_{\nu}=0~\rightarrow E= p_{r} v(r)\pm p_{r}~.
\end{equation}
Here $g^{\mu \nu}$ is the contravariant metric,  $p_{r}$ is the radial momentum, and $E=p_{0}$.

The exponent of the imaginary part of the action along the tunneling trajectory, Im $\int p_{r}(r, E) dr$, gives the probability of the tunneling process, where the trajectory $p_{r}(r, E)$ is,
\begin{eqnarray}
 && p_{r}(r,E)= \frac{E}{v(r)-1}\nonumber\\
 &&=\frac{E}{\sqrt{\frac{2 M}{r}+\frac{r^2}{l^2}}-1}\nonumber\\
 &&= -\frac{E \left(\sqrt{\frac{2 M}{r}+\frac{r^2}{l^2}}+1\right)}{1-\frac{2 M}{r}-\frac{r^2}{l^2}}~.\label{eq10}
\end{eqnarray}

In terms of the horizons and surface gravities, $\left(1-\frac{2 M}{r}-\frac{r^2}{l^2}\right)^{-1}$ can be expressed as,
\begin{eqnarray}
&&\left(1-\frac{2 M}{r}-\frac{r^2}{l^2}\right)^{-1}\nonumber\\&&=\frac{\alpha}{2 \kappa_{h}(r-r_{h})}+\frac{\alpha}{2 \kappa_{c}(r-r_{c})}-\frac{\alpha}{2 \kappa_{0}(r-r_{0})}~.\nonumber\\
\end{eqnarray}
Here $r_{0}=-(r_{c}+r_{h})$ being
negative, is unphysical and surface gravity associated with $r_{0}$ is defined as $\kappa_0=\frac{1}{2}\left|\frac{\partial f(r)}{\partial r}\right|_{r=r_0}$.
So, Im $\int p_{r}(r, E) dr$ can also be expressed as,
\begin{eqnarray}
&&Im \int p_{r}(r,E) dr\nonumber\\ 
&=& - Im \int E \left(\sqrt{\frac{2 M}{r}+\frac{r^2}{l^2}}+1\right)\times \left(\frac{\alpha}{2 \kappa_{h}(r-r_{h})}\right.\nonumber\\&&\left.+\frac{\alpha}{2 \kappa_{c}(r-r_{c})} -\frac{\alpha}{2 \kappa_{0}(r-r_{0})}\right)~.\label{eq12}
\end{eqnarray}
From eq. (\ref{eq12}), it can be easily shown that the contribution of two horizons for this spacetime (\ref{eq1}) gives the following probability of the Hawking radiation as,
\begin{equation}\label{eq13}
P=exp\left(- \frac{2 \pi \alpha E}{\kappa_{eff}}\right),
\end{equation}
where $\kappa_{eff}=\left(\frac{1}{\kappa_{h}}+\frac{1}{\kappa_{c}}\right)^{-1}$. This corresponds to thermal radiation characterized by the Hawking temperature, where the Hawking temperature is given by,
\begin{equation}\label{eq14}
T_{H}=\frac{\kappa_{eff}}{2 \pi \alpha}= \frac{\kappa_{c}\kappa_{h}}{2 \pi \alpha(\kappa_{c}+\kappa_{h})}~.
\end{equation}
This result is precisely matched with the result in the paper \cite{Shankaranarayanan:2003ya}.
Eq. (\ref{eq14}) implies that a global temperature can exist for the Schwarzschild-de Sitter spacetime due to the presence of the two horizons.

\subsection{Interpretation in terms of the Hawking temperatures at two horizons of the Schwarzschild-de Sitter spacetime}

Eq. (\ref{eq14}) can be expressed in terms of the Hawking temperature at two horizons of Schwarzschild-de Sitter spacetime, where the probability of the Hawking radiation eq. (\ref{eq13}) can be written as
\begin{equation}
P=P_{c} P_{h}= exp\left(-\frac{E}{T_{c}}\right) \times exp\left(-\frac{E}{T_{h}}\right)~.
\end{equation}
Here $T_{h}$ and $T_{c}$ are the conventional Hawking temperatures at the black hole horizon and cosmological horizon, respectively. The Hawking temperatures associated with these two horizons are given below,
\begin{align}
T_{h}=\frac{\kappa_{h}}{2 \pi \alpha}~,\\
T_{c}=\frac{\kappa_{c}}{2 \pi \alpha}~.
\end{align}
The temperature $T_{c}$ determines the rate of the tunneling from the region $r_{h}<r< r_{c}$ to $r>r_{c}$, while the $T_{h}$ determines the occupation number of these particle near the black hole horizon. The presence of the two horizons leads to the final probability of the Hawking radiation as,
\begin{equation} \label{eq18}
P=P_{c}P_{h}= exp\left(-\frac{E}{T_{H}}\right).
\end{equation}
Eq. (\ref{eq18}) implies that the
global temperature does not coincide with the conventional Hawking temperature related to the outer cosmological horizon.

\subsection{Entropy for Schwarzschild-de Sitter spacetime by using the approach based on the singular coordinate transformations}

The coordinate transformations in Eq. (\ref{eq6}) have two singularities. So the macroscopic quantum tunneling \cite{Volovik:2021upi, Volovik:2021iim} from the PG Schwarzschild-de Sitter (Sds) to its static partner with the same energy $E$ is determined by the following exponent,

\begin{eqnarray}
&& P_{Sds \rightarrow static} = exp\left(- 2 Im \int E d\tilde{t}\right)\nonumber\\
&=& exp \left(-2 Im \int E(dt + f(r)dr)\right)\nonumber\\
&=& exp \left(-2 E Im \int f(r)dr\right)\nonumber\\
&=& exp(-2 E Im  \int  \left(\sqrt{\frac{2 M}{r}+\frac{r^2}{l^2}}\right)\times\nonumber\\&&\left(\frac{\alpha}{2 \kappa_{h}(r-r_{h})}+\frac{\alpha}{2 \kappa_{c}(r-r_{c})}-\frac{\alpha}{2 \kappa_{0}(r-r_{0})}\right)\nonumber\\
&=& exp\left(- \frac{  \pi \alpha E}{\kappa_{eff}}\right)~.\label{eq19}
\end{eqnarray}
The Eq. (\ref{eq19}) gives the entropy of the Sds spacetime as,
\begin{equation} \label{eq20}
S_{Sds}= \frac{\pi}{\kappa^2_{eff}}~,
\end{equation}
where we take $E=1/\alpha\kappa_{eff}$. Here we assume that the relation between the entropy and surface gravity for single horizon spacetime holds true for multi-horizon spacetime. From eq. (\ref{eq20}), it can be shown that the entropy for Schwarzschild-de Sitter spacetime can  be expressed as $S_{Sds}=(\sqrt{S_{S}}+\sqrt{S_{ds}})^2$. Here $S_{S} =\frac{\pi}{\kappa^2_{h}}$, and $S_{ds} =\frac{\pi}{\kappa^2_{c}}$ are the entropy associated with black hole horizon, and the entropy associated with cosmological horizon, respectively.
So the total entropy for Schwarzschild-de Sitter spacetime can not be determined only by the outer cosmological horizon. The correlations between the two horizons determine it.
This result exactly matches with the previous conclusion in the paper \cite{Shankaranarayanan:2003ya}. 

\section{HAWKING RADIATION FROM THREE
HORIZONS OF the Reissner-Nordstrom-de Sitter spacetime}\label{Reissner-Nordstrom-de Sitter}

The line element for the Reissner-Nordstrom-de Sitter spacetime in the spherically symmetric coordinate is given by \cite{Li:2021axp, Zhang:2016nws, Hollands:2019whz, Guo:2005hw, Ahmed:2016lou},
\begin{equation}\label{eq23}
ds^2=-f(r)dt^2+f(r)^{-1} dr^2 +r^2 d \Omega^2~,
\end{equation}
where $f(r)=\left(1-\frac{2 M}{r}+\frac{Q^2}{r^2}-\frac{\Lambda r^2}{3}\right)$. Here $M$ is the mass of the black hole $Q$ is the charge of the black hole, and $\Lambda$ is the positive cosmological constant. There are three  horizons, $r_{+}$ $r_{-}$ and $r_{c}$, for the line element (\ref{eq23}). Here $r_{+}$ is the event horizon, $r_{-}$ is the Cauchy horizon and $r_{c}$ is the outer cosmological horizon. The surface gravities at the horizons are given by, 
\begin{align}
\kappa_{+}= \left|\frac{M}{r^2_+}- \frac{Q^2}{r^3_{+}}-\frac{\Lambda r_+}{3}\right|~,\\
\kappa_{-}= -\left|\frac{M}{r^2_-}- \frac{Q^2}{r^3_{-}}-\frac{\Lambda r_-}{3}\right|~,\\
\kappa_{c}= \left|\frac{M}{r^2_c}- \frac{Q^2}{r^3_{c}}-\frac{\Lambda r_c}{3}\right|~,
\end{align}
where $\kappa_{+}$, $\kappa_{-}$ and $\kappa_{c}$ are the surface gravities of the event horizon, Cauchy horizon, and cosmological horizon, respectively. We get the PG metric for Reissner-Nordstrom-de Sitter spacetime using the following coordinate transformations,
\begin{equation}\label{eq24}
 d\tilde{t}\rightarrow dt \pm f dr,~~f=\sqrt{\frac{\left(\frac{2M}{r}+\frac{\Lambda r^2}{3}\right)}{ \left(1+\frac{Q^2}{r^2}\right)}}\frac{1}{\left(1-\frac{2 M}{r}+\frac{Q^2}{r^2}-\frac{\Lambda r^2}{3}\right)}~,
\end{equation}
where the PG metric is given by,
\begin{eqnarray}\label{eq21}
ds^2&=& g_{\mu \nu} dx^{\mu} dx^{\nu}\nonumber\\&=&-\left(1+\frac{Q^2}{r^2}\right) dt^2+\frac{1}{\left(1+\frac{Q^2}{r^2}\right)}(dr\pm v dt)^2+r^2 d\Omega^2~. \nonumber\\
\end{eqnarray}
Here the shift velocity is,
\begin{equation}
 v^2=\left(\frac{2 M}{r}+\frac{\Lambda r^2}{3}\right)\left(1+\frac{Q^2}{r^2}\right)~.
\end{equation}
The tunneling trajectory for a massless particle can be found as,
\begin{equation}
 g^{\mu\nu}p_{\mu}p_{\nu}=0~\rightarrow E= p_{r} v(r)\pm p_{r} \left(1+\frac{Q^2}{r^2}\right)~.
\end{equation}
Here $g^{\mu \nu}$ is the contravariant metric,  $p_{r}$ is the radial momentum, and $E=p_{0}$.

The exponent of the imaginary part of the action along the tunneling trajectory, Im $\int p_{r}(r, E) dr$, gives the probability of the tunneling process, where the trajectory $p_{r}(r, E)$ is given by,
\begin{eqnarray}
 p_{r}(r,E)&=& \frac{E}{v(r)-\left(1+\frac{Q^2}{r^2}\right)}\nonumber\\
 &=&\frac{E}{\sqrt{\left(\frac{2 M}{r}+\frac{\Lambda r^2}{3}\right)\left(1+\frac{Q^2}{r^2}\right)}-\left(1+\frac{Q^2}{r^2}\right)}\nonumber\\
 &=& -\frac{E \left(\sqrt{\left(\frac{2 M}{r}+\frac{\Lambda r^2}{3}\right)}+\sqrt{\left(1+\frac{Q^2}{r^2}\right)}\right)}{\sqrt{\left(1+\frac{Q^2}{r^2}\right)} \left(1-\frac{2 M}{r}+\frac{Q^2}{r^2}-\frac{\Lambda r^2}{3}\right)}~.\label{eq31}
\end{eqnarray}

In terms of the horizons and surface gravities, $\left(1-\frac{2 M}{r}+\frac{Q^2}{r^2}-\frac{\Lambda r^2}{3}\right)^{-1}$ can be expressed as,
\begin{eqnarray}
&&\left(1-\frac{2 M}{r}+\frac{Q^2}{r^2}-\frac{\Lambda r^2}{3}\right)^{-1}=\frac{1}{2 \kappa_{+}(r-r_{+})}\nonumber\\&&+\frac{1}{2 \kappa_{-}(r-r_{-})} +\frac{1}{2 \kappa_{c}(r-r_{c})}-\frac{1}{2 \kappa_{0}(r-r_{0})}~.\nonumber\\
\end{eqnarray}
Here $r_{0}=-(r_{+}+r_{-}+r_{c})$ being
negative, is unphysical and surface gravity associated with $r_{0}$ is defined as $\kappa_0=\frac{1}{2}\left|\frac{\partial f(r)}{\partial r}\right|_{r=r_0}$.
So, Im $\int p_{r}(r, E) dr$ can also be expressed as,
\begin{eqnarray}
&&Im \int p_{r}(r,E) dr\nonumber\\ 
&=& - Im \int \frac{E \left(\sqrt{\left(\frac{2 M}{r}+\frac{\Lambda r^2}{3}\right)}+\sqrt{\left(1+\frac{Q^2}{r^2}\right)}\right)}{\sqrt{\left(1+\frac{Q^2}{r^2}\right)}}\times \nonumber\\ &&  \left(\frac{1}{2 \kappa_{+}(r-r_{+})}+\frac{1}{2 \kappa_{-}(r-r_{-})}+\frac{1}{2 \kappa_{c}(r-r_{c})}\right.\nonumber\\&&\left.-\frac{1}{2 \kappa_{0}(r-r_{0})}\right)~.\nonumber\\ \label{eq33}
\end{eqnarray}
From eq. (\ref{eq33}), it can be easily shown that the contribution of three horizons gives the following probability of the Hawking radiation as,
\begin{equation}\label{eq32}
P=exp\left(- \frac{2 \pi E}{\kappa_{eff}}\right),
\end{equation}
where $\kappa_{eff}=\left(\frac{1}{\kappa_{+}}+\frac{1}{\kappa_{-}}+\frac{1}{\kappa_{c}}\right)^{-1}$. This corresponds to the thermal radiation which is characterized by the Hawking temperature, where the Hawking temperature is given by,

\begin{equation}\label{eq31}
T_{H}=\frac{\kappa_{eff}}{2 \pi}~.
\end{equation}
Eq. (\ref{eq31}) implies that a global temperature can exist for the Reissner-Nordstrom-de Sitter Spacetime due to the presence of three horizons.

\subsection{Interpretation in terms of the Hawking
temperatures at three horizons of the Reissner-Nordstrom de-Sitter spacetime}

Eq. (\ref{eq31}) can be interpreted in terms of the Hawking temperature at three horizons of the Reissner-Nordstrom-de Sitter Spacetime, where the probability of the Hawking radiation eq. (\ref{eq32}) can be written as
\begin{equation}
P=P_{+} P_{-} P_{c}= exp\left(-\frac{E}{T_{+}}\right) exp\left(-\frac{E}{T_{-}}\right) exp\left(-\frac{E}{T_{c}}\right)~.
\end{equation}
Here $T_{+}$, $T_{-}$ and $T_{c}$ are the conventional Hawking temperatures at the event horizon, Cauchy horizon, and cosmological horizon, respectively. The Hawking temperatures associated with these three horizons are given below,
\begin{align}
T_{+}=\frac{\kappa_{+}}{2 \pi}~,\\
T_{-}=\frac{\kappa_{-}}{2 \pi}~,\\
T_{c}=\frac{\kappa_{c}}{2 \pi}~.
\end{align}
The temperature $T_{c}$ determines the rate of the tunneling from the region $r_{+}<r< r_{c}$ to $r>r_{c}$, $T_{+}$ determines the rate of the tunneling from the region $r_{+}<r< r_{-}$ to $r>r_{+}$, while the $T_{-}$ determines the occupation number of these particle near the inner Cauchy horizon. The presence of the three horizons leads to the final probability of the Hawking radiation as,
\begin{equation} \label{eq40}
P=P_{+}P_{-} P_{c}= exp\left(-\frac{E}{T_{H}}\right).
\end{equation}
Eq. (\ref{eq40}) implies that the
global temperature does not coincide with the conventional Hawking temperature related to the outer cosmological horizon.

\subsection{Entropy for Reissner-Nordstrom-de Sitter spacetime by using the approach based on the singular coordinate transformations}

The coordinates transformations in Eq. (\ref{eq24}) have three singularities. So the macroscopic quantum tunneling \cite{Volovik:2021upi, Volovik:2021iim}  from the PG Reissner-Nordstrom-de Sitter (RNds) to its static partner is determined by the following exponent,
\begin{eqnarray}
&& P_{RNds \rightarrow static} = exp\left(- 2 Im \int E d\tilde{t}\right)\nonumber\\
&=& exp \left(-2 Im \int E(dt + f(r)dr)\right)\nonumber\\
&=& exp \left(-2 E Im \int f(r)dr\right)\nonumber\\
&=& exp(-2 E Im  \displaystyle \int  \left(\sqrt{\frac{\frac{2 M}{r}+\frac{\Lambda r^2}{3}}{1+\frac{Q^2}{r^2}}}\right)\times\left(\frac{1}{2 \kappa_{+}(r-r_{+})}\right.\nonumber\\&&\left.+\frac{1}{2 \kappa_{-}(r-r_{-})}+\frac{1}{2 \kappa_{c}(r-r_{c})}-\frac{1}{2 \kappa_{0}(r-r_{0})}\right)~.\nonumber\\
&=& exp\left(- \frac{  \pi E}{\kappa_{eff}}\right)~.\label{eq41}
\end{eqnarray}
The Eq. (\ref{eq41}) gives the entropy of the RNds spacetime as,
\begin{equation}\label{eq42}
S_{RNds}= \frac{\pi}{\kappa^2_{eff}}~,
\end{equation}
where we take $E=1/\kappa_{eff}$.  Here we assume that the relation between the entropy and surface gravity for single horizon spacetime holds true for multi-horizon spacetime. From eq. (\ref{eq42}), it is shown that the entropy for  Reissner-Nordstrom-de Sitter spacetime can  be expressed as $S_{RNds}=(\sqrt{S_{+}}+\sqrt{S_{-}}+\sqrt{S_{ds}})^2$. Here  $S_{+} =\frac{\pi}{\kappa^2_{+}}$, $S_{-}=\frac{\pi}{\kappa^2_{-}}$, and $S_{ds} =\frac{\pi}{\kappa^2_{c}}$ are the entropy associated with the event horizon, the entropy associated with the Cauchy horizon, and the entropy associated with the cosmological horizon, respectively.
So the total entropy for Reissner-Nordstrom-de Sitter spacetime is not determined only by the outer cosmological horizon. The correlations between the three horizons determine it.

\section{HAWKING RADIATION FROM TWO
HORIZONS OF Rotating BTZ black hole}\label{BTZ}

The line element for a rotating BTZ black hole is given by \cite{Banados:1992wn,Banados:1992gq,Dias:2019ery, Chaturvedi:2013ova, Kajuri:2020bvi,Fathi:2021eig},
\begin{equation}\label{eq45}
ds^2=-f(r)dt^2+f(r)^{-1} dr^2 +r^2 \left(d \phi-\frac{J}{2r^2}\right)^2~,
\end{equation}
where $f(r)=\left(-M+\frac{r^2}{l^2}+\frac{J^2}{4 r^2}\right)=\frac{(r^2-r^2_{+})(r^2-r^2_{-})}{l^2 r^2}$.  Here $M$ is the mass of the BTZ black hole, $l^2$ is related to the negative cosmological constant, $J$ is the angular momentum and $r_{\pm}=l~\left(\frac{M}{2}\left(1\pm\sqrt{1-(\frac{J}{M l})^2}\right)\right)^{1/2}$ is the inner Cauchy horizon and outer event horizon, respectively. We get the PG metric for rotating BTZ black hole using following coordinate transformations as,

\begin{equation}\label{eq46}
 d\tilde{t}\rightarrow dt \pm f dr,~~f=\sqrt{\frac{M}{\frac{r^2}{l^2}+\frac{J^2}{4 r^2}}}.\frac{1}{\left(-M+\frac{r^2}{l^2}+\frac{J^2}{4 r^2}\right)}~.
\end{equation}
where the PG metric is given by,
\begin{eqnarray}\label{eq47}
&& ds^2 = g_{\mu \nu} dx^{\mu} dx^{\nu}\nonumber\\&&=-\left(\frac{r^2}{l^2}+\frac{J^2}{4 r^2}\right) dt^2+\frac{1}{\left(\frac{r^2}{l^2}+\frac{J^2}{4 r^2}\right)}(dr\pm v dt)^2\nonumber\\&&+r^2 \left(d \phi-\frac{J}{2r^2}\right)^2~.\nonumber\\ 
\end{eqnarray}
Here the shift velocity is,
\begin{equation}
 v^2=M \left(\frac{r^2}{l^2}+\frac{J^2}{4 r^2}\right)~.
\end{equation}
The tunneling trajectory for a massless particle can be found as,
\begin{equation}
 g^{\mu\nu}p_{\mu}p_{\nu}=0~\rightarrow E= p_{r} v(r)\pm p_{r} \left(\frac{r^2}{l^2}+\frac{J^2}{4 r^2}\right)~.
\end{equation}

Here $g^{\mu \nu}$ is the contravariant metric,  $p_{r}$ is the radial momentum, and $E=p_{0}$.

The exponent of the imaginary part of the action along the tunneling trajectory, Im $\int p_{r}(r, E) dr$, gives the probability of the tunneling process, where the trajectory $p_{r}(r, E)$ is given by
\begin{eqnarray}
 && p_{r}(r,E)= \frac{E}{v(r)-\left(\frac{r^2}{l^2}+\frac{J^2}{4 r^2}\right)}\nonumber\\
 &&=\frac{E}{\sqrt{M \left(\frac{r^2}{l^2}+\frac{J^2}{4 r^2}\right)}-\left(\frac{r^2}{l^2}+\frac{J^2}{4 r^2}\right)}\nonumber\\
 &&= -\frac{E \left(\sqrt{M}+\sqrt{ \left(\frac{r^2}{l^2}+\frac{J^2}{4 r^2}\right)}\right)}{\sqrt{ \left(\frac{r^2}{l^2}+\frac{J^2}{4 r^2}\right)}\left(-M+\frac{r^2}{l^2}+\frac{J^2}{4 r^2}\right)}~.\label{eq50}
\end{eqnarray}
As $\left(-M+\frac{r^2}{l^2}+\frac{J^2}{4 r^2}\right)^{-1}$ can be expressed in terms of horizons as,
\begin{eqnarray}
\left(-M+\frac{r^2}{l^2}+\frac{J^2}{4 r^2}\right)^{-1}=\frac{l^2 r^2}{(r^2-r^2_{+})(r^2-r^2_{-})}~.\nonumber\\
\end{eqnarray}
So, Im $\int p_{r}(r,E) dr$ can also be expressed as,
\begin{eqnarray}
&&Im \int p_{r}(r,E) dr \nonumber\\&&= - Im \int \frac{E \left(\sqrt{M}+\sqrt{ \left(\frac{r^2}{l^2}+\frac{J^2}{4 r^2}\right)}\right)}{\sqrt{ \left(\frac{r^2}{l^2}+\frac{J^2}{4 r^2}\right)}}\frac{l^2 r^2}{(r^2-r^2_{+})(r^2-r^2_{-})}~.\nonumber\\\label{eq52}
\end{eqnarray}
From eq. (\ref{eq52}), it can be easily shown that the contribution of two horizons gives the following probability of the Hawking radiation as,
\begin{equation}\label{eq50}
P=exp\left(- \frac{2 \pi l^2~E}{(r_{+}+r_{-})}\right)~.
\end{equation}
 This corresponds to the thermal radiation, which is characterized by the Hawking temperature, where the Hawking temperature is given by,
\begin{equation}\label{eq54}
T_{H}=\frac{(r_{+}+r_{-})}{2 \pi l^2}~.
\end{equation}
Eq. (\ref{eq54}) implies that a global temperature can exist for the rotating BTZ black hole due to the presence of two horizons.

\subsection{Interpretation in terms of the Hawking
temperatures at two horizons of the
rotating BTZ black hole}

 Eq. (\ref{eq54}) can be interpreted in terms of the  Hawking temperature at two horizons of the rotating BTZ black hole, where the probability of the Hawking radiation eq. (\ref{eq50}) can be written as
\begin{equation}
P=P_{+} P_{-}= exp\left(-\frac{E}{T_{+}}\right) \times exp\left(-\frac{E}{T_{-}}\right)~.
\end{equation}
Here $T_{+}$ and $T_{-}$ are the conventional  Hawking temperatures that correspond to the event horizon and Cauchy horizon, respectively, which are given by,
\begin{align}
T_{+}=\frac{\kappa_+}{2 \pi}=\frac{1}{4 \pi}\frac{\partial f(r)}{\partial r}|_{r=r_+}=\frac{(r^2_{+}-r^2_{-})}{2 \pi l^2 r_{+}}~,\\
T_{-}=\frac{\kappa_-}{2 \pi}=\frac{1}{4 \pi}\frac{\partial f(r)}{\partial r}|_{r=r_-}=-\frac{(r^2_{+}-r^2_{-})}{2 \pi l^2 r_{-}}~.
\end{align}
Here $\kappa_+$ and $\kappa_-$ are the surface gravities for the event horizon and the Cauchy horizon, respectively.
The temperature $T_{+}$ determines the rate of the tunneling from the region $r_{-}<r< r_{+}$ to $r>r_{+}$, while the $T_{-}$ determines the occupation number of these particles near the inner Cauchy horizon. The presence of the two horizons leads to the final probability of the Hawking radiation as,
\begin{equation}\label{eq58}
P=P_{+}P_{-}= exp\left(-\frac{E}{T_{H}}\right).
\end{equation}
Eq. (\ref{eq58}) implies that the
global temperature does not coincide with the conventional Hawking temperature related to the outer event horizon.

\subsection{Entropy for rotating BTZ black hole by using the approach based on the singular
coordinate transformations}

The coordinate transformations in Eq. (\ref{eq46}) have two singularities. So the macroscopic quantum tunneling \cite{Volovik:2021upi, Volovik:2021iim}  from the PG rotating BTZ to its static partner is determined by the following exponent,
\begin{eqnarray}
&& P_{BTZ \rightarrow static} = exp\left(- 2 Im \int E d\tilde{t}\right)\nonumber\\
&=& exp \left(-2 Im \int E(dt + f(r)dr)\right)\nonumber\\
&=& exp \left(-2 E Im \int f(r)dr\right)\nonumber\\
&=& exp \left(-2 E Im  \int \sqrt{\frac{M}{\frac{r^2}{l^2}+\frac{J^2}{4 r^2}}}\frac{l^2 r^2}{(r^2-r^2_{+})(r^2-r^2_{-})}\right)\nonumber\\
&=& exp \left(-\frac{\pi E l^2}{(r_{+}+r_{-})}\right)~.\label{eq59}  
\end{eqnarray}
The Eq. (\ref{eq59}) gives the entropy of the rotating BTZ black hole as,
\begin{equation}\label{eq60}
S_{BTZ}= \frac{\pi l^4}{(r_{+}+r_{-})^2}~.
\end{equation}
Here we take $E=1/\kappa_{eff}$ , where $\kappa_{eff}=\left(\frac{1}{\kappa_{+}}+\frac{1}{\kappa_{-}}\right)^{-1}=(r_{+}+r_{-})/l^2$.  Here we assume that the relation between the entropy and surface gravity for single horizon spacetime holds true for multi-horizon spacetime. From eq. (\ref{eq60}), it is shown that the entropy for  can  be expressed as $S_{BTZ}=(\sqrt{S_{+}}+\sqrt{S_{-}})^2$. Here $S_{+} =\frac{\pi}{\kappa^2_{+}}$, and $S_{-}=\frac{\pi}{\kappa^2_{-}}$ are the entropy associated with the event horizon, and the entropy associated with the Cauchy horizon, respectively.
So the total entropy for a rotating BTZ black hole is not determined only by the outer event horizon. The correlations between the two horizons determine it.

\section{DISCUSSIONS}

In this paper, we have calculated thermal radiation and the entropy for the multi-horizon spacetime. We have shown that a global temperature can exist due to the presence of the multi horizons. For Schwarzschild-de Sitter spacetime, the global temperature does not coincide with the conventional Hawking temperature related to the outer cosmological horizon. Also, the total entropy for this spacetime can not be determined only by the outer cosmological horizon. The correlations between the cosmological horizon and event horizon determine it. For Reissner-Nordstrom-de Sitter spacetime, the global temperature does not coincide with the conventional Hawking temperature related to the outer cosmological horizon. Also, the total entropy for this spacetime can not be determined only by the outer cosmological horizon. The correlations between the cosmological horizon, event horizon, and Cauchy horizon determine it.
Similarly, for rotating BTZ black holes,  the global temperature does not coincide with the conventional Hawking temperature related to the outer event horizon. The total entropy for this spacetime can also not be determined only by the outer event horizon. The correlations between the event horizon and the Cauchy horizon determine it. 

It would be quite interesting  to apply the above technique for other spacetimes such as Kerr-de Sitter spacetime \cite{Akcay:2010vt,LI2017211,10.1143/PTP.100.491}, Kerr–Newman–de Sitter spacetimes \cite{Franzen:2020gke,Gwak:2018tmy,PhysRevD.58.084003}, charged BTZ black hole \cite{Hendi:2015wxa,Hendi:2020yah,Tang:2016vmu} and black holes in Horava gravity \cite{Janiszewski:2014iaa, Davison:2016auk} for defining a global temperature and entropy for these spacetimes.

Recently, it has been shown that the Hawking radiation from the
Lemaitre-Tolman–Bondi model with a positive cosmological constant was recovered from exact solutions to the Wheeler-Dewitt equation and the momentum constraints \cite{Franzen:2009ev}. The exact calculation would provide the temperature of the thermal radiation which does not coincide with the conventional Hawking temperature related to the outer horizon. Thus we hope that our results can also be derived from the exact solutions to the Wheeler-Dewitt equation and the momentum constraints in the spacetime considered here. This we leave for the future.

For the pure de Sitter (\emph{i.e.} in spacetime without the black hole), there are still controversies related to the Hawking radiation from the cosmological horizon \cite{Kamenshchik:2021tjh}. In particular, it is not excluded that the universe expansion can exactly compensate the Hawking radiation from the cosmological horizon due to the high symmetry of the de Sitter spacetime. This means that the Hawking radiation from the cosmological horizon does not lead to dissipation, contrary to the Hawking radiation from the event horizon. In the spacetime considered here, the de Sitter symmetry is violated.  This violation of high symmetry may allow avoiding the controversy, \emph{i.e.}, the cosmological horizon dissipates. It would be quite interesting to look into this problem with special consideration. This is because the de Sitter problem is of fundamental importance.
\bigskip
\begin{acknowledgments}

I thank Sumanta Chakraborty, Habib Ahammad Mondal, and Sudip Mandal for many useful discussions. I also thank the Saha Institute of Nuclear Physics (SINP) Kolkata for financial support.
\end{acknowledgments}

\section*{Data availability}

This manuscript has no associated data or
the data will not be deposited. [Authors’ comment: The work is purely
analytical. We have not utilized any particular external set of data to
produce the results.]

%\bibliography{bibtexfile}


\end{document}